\documentclass{acmart}
\pdfoutput=1
\usepackage[utf8]{inputenc}

\usepackage{wrapfig}
\usepackage{natbib}
\usepackage{graphicx}

\begin{document}

\title{CoRank: A clustering cum graph ranking approach for extractive summarization}

\author{Mohd Khizir Siddiqui}
\authornote{Both authors contributed equally to this research.}
\email{f20180439@goa.bits-pilani.ac.in}
\orcid{0000-0002-7471-1118}
\affiliation{%
  \institution{Birla Institute of Technology and Science, Pilani}
  \country{India}
}

\author{Amreen Ahmad}
\authornotemark[1]
\affiliation{%
  \institution{Jamia Millia Islamia}
  \city{New Delhi}
  \country{India}}
\email{amreen.ahmad10@gmail.com}

\author{Om Pal}
\email{ompal.cdac@gmail.com }
\affiliation{%
  \institution{MeitY}
  \country{India}
}

\author{Tanvir Ahmad}
\affiliation{%
	\institution{Jamia Millia Islamia}
	\city{New Delhi}
	\country{India}}
\email{tahmad2@jmi.ac.in}

\renewcommand{\shortauthors}{Siddiqui et al.}

\begin{abstract}
  On-line information has increased tremendously in today’s age of Internet. As a result, the need has arose to extract relevant content from the plethora of available information. Researchers are widely using automatic text summarization techniques for extracting useful and relevant information from voluminous available information, it also enables users to obtain valuable knowledge in a limited period of time with minimal effort. The summary obtained from the automatic text summarization often faces the issues of diversity and information coverage. Promising results are obtained for  automatic text summarization by the introduction of new techniques based on graph ranking of sentences, clustering, and optimization. This research work proposes CoRank, a two-stage sentence selection model involving clustering and then ranking of sentences.  The initial stage involves clustering of sentences using a novel clustering algorithm, and later selection of salient sentences using CoRank algorithm.Te approach aims to cover two objectives: maximum coverage and diversity, which is achieved by the extraction of main topics and sub-topics from the original text.  
  The performance of the CoRank is validated on DUC2001 and DUC 2002 data sets.
\end{abstract}


\maketitle

\section{Introduction}
Recently, multi-document summarization and document clustering have gained a lot of attention for analyzing textual information. The aim of document clustering is to divide a set of documents into distinct classes called clusters, where similar documents are occurring in same cluster and dissimilar documents \cite{Duda2001} occurring in different clusters. Another efficient technique for extracting relevant content from huge volumnious data is multi-document summarization, that aims to create reduced summary while retaining the gist of the documents \cite{Mani, RICARDO}. Both these techniques find a range of applications in information retrieval and management, apart from retrieving relevant content from documents. Results of web search \cite{ZAMIR} can be organized and presented in an efficient way using document clustering. Apart from it, multi-document summarization finds its use in creation of snippet on the Web, that can be further used for future purposes \cite{TURPIN}. 

Usually researchers while performing document clustering consider the set of documents as document term matrix where documents are represented by rows and terms are represented by columns. Several conventional clustering algorithms exist for grouping the similar documents. But these algorithms were unable to correctly interpret the document cluster. In recent years, some works \cite{DHILLON} have focused on capturing the dual knowledge between documents and terms and performing clustering at the same time. But this framework has limitation, since every document cluster is represented by a set of representative words, and these words cannot give a true interpretation of clusters as they lack contextual and semantic information. Another way could be selection of salient sentences from the cluster.

Multi-document summarization(MDS) aims at producing a condensed version of the document while retaining the girth of the document.  Based on the output type, MDS can be categorized into abstractive or extractive.  In extractive summarization, representative sentences are selected as summary from the original source document. This approach uses some pre-defined methodolgy to compute sentence scores and based on that salient sentences are seelcted. Whereas in abstractive summarization, summary is composed of salient sentences that represent the gist of the document but are not part of original document. Some techniques such as reformulation, sentence compression, and information fusion is involved in abstractive summarization. The summary obtained from abstractive summarization are compact since it uses deep learning techniques. This research work concentrates on extractive summarization since it is more practical and feasible.  In MDS, the set of documents are represented as sentence-term matrix where rows and columns are represented as sentences and terms. Numerous clustering approaches \cite {JING, KNIGHT} have been proposed for extractive summarization where the first step is cluster generation and then identification of salient sentences from these clusters. The limitation of such approach is that they consider sentences as independent units and contextual dependency among sentences are ignored. However, the mutual influence of sentences occurring within same cluster should be considered for correct interpretation of cluster. 

The focus of the proposed methodology in the research paper is to convert a document into an appropriate graph structure, cluster it into overlapping communities, and extract out most important sentences of the document. The novelty of the proposed research work is highlighted below:

\subsection{Contributions}
\begin{itemize}
    \item The given textual document is converted into a graph. A novel algorithm is developed to detect overlapping communities in a weighted network with feasible computational requirements.
    \item A node can lie in several of the overlapping communities. This invalidates the assumption of the influential nodes algorithm on unweighted, distinct communities. We propose an inexpensive algorithm on top of the community detection to list out important nodes of the underlying graph. 
    \item Larger subtopics in a document carry more importance than the smaller subtopics. The proposed CoRank method picks sentences depending on the weightage of subtopic.
    \item We conduct experiments on the standard DUC-2001 and DUC-2002 data-set to validate the performance of proposed CoRank method.
\end{itemize}

The further sections are divided as follows. Section \ref{sec:rel_work} introduces overview of the work done in this area and connected relevant fields followed by a brief introduction of the problem statement given in Section 3. Section \ref{sec:method} discusses a step-by-step procedure for the proposed methodology. Section \ref{sec:example} discusses about the explanation of the proposed methodolgy using a toy network example. Section 6 describes in detail the analysis of the experiments followed by conclusion in   Section \ref{sec:conclusion}.

\section{Related Work} \label{sec:rel_work}

Extractive text summarization has been in active developments in recent years, numerous methods have been proposed to solve the problem. The key idea lies in developing an efficient scoring method for the sentences in the document. Many methods apply topic-wise clustering on the document and identifying key individual sentences with respect to the topics. Some other approaches revolve around using evolutionary algorithms \cite{Mendoza2014ExtractiveSS} or machine learning techniques \cite{Ozsoy2011} \cite{Jang2021}.  

\subsection{Graph Based Approaches}

The method proposed in \cite{CAI20113816} simultaneously clusters the sentences and scores them for a ranking. The focus of work in \cite{Alguliyev2017AMF} is to cluster the sentences and perform their selection as a solution to an optimization problem. \cite{CAI2014764} targets both diversity and coverage of the summary using an integrated clustering based technique. The idea in \cite{10.1145/2513563} uses co-clustering method on words and sentences individually to perform a topic based summarization. The framework introduced allows words to have an explicit decision in sentence selection to squeeze out better performance. \cite{Mei2012} proposes a fuzzy c-medoid based clustering approach to produce cluster of sentences similar to a subtopic of the topic. A tool named Compendium is proposed in \cite{lloret2013}, which combines textual entailment, statistical and cognition based techniques to remove redundant information and find relevant content in summary. The work in \cite{Luo2013} focuses on probabilistic modelling topic relevance and coverage in summarization. In \cite{Yang2008}, the authors use fractal theory to infer the interplay of sentences and perform the summarization.

The work in \cite{Ferreira2014} uses a graph based approach to cluster the sentences. Document is modelled as a graph and then the different methods are used to rank the sentences (modelled as nodes). \cite{li-li-2014-query} propose a semi-supervised clustering method on the graphs combined with topic modelling. In \cite{Balaji2014}, they use the ideas of graph matching to improve upon the results. \cite{Wan2007} proposes \textit{Collabsum}, which exploits information from multiple documents by clustering them and extracting the mutual influence to summarise a single document. This methodology incorporates both the intra-document and inter-document relationships. \cite{10.1016/j.knosys.2012.05.017} models the summarization as a modified p-median problem. The work in \cite{Hingu2015}, uses external knowledge from Wikipedia to enhance performance on the existing graph based methods. In \cite{Erkan2004}, a novel \textit{LexRank} method is introduced, it determines the salience using eigenvector centrality in graphical representation. The work in \cite{wan-2010-towards} models the documents as graph and exploits mutual information between documents to generate summary.

A separate line of work has evolved in community detection and influential node identification. \cite{Yang2018} introduces \textit{TOPSIS} to identify influential nodes by considering it as a multi-attribute decision-making problem. \cite{Ding2016} proposes \textit{NDOCD} where links are iteratively removed to reduce the graph into clusters. In \cite{Zhang2020}, the method proposed network embedding is used to decompose the network into communities and then nodes are chosen to maximize influence. \cite{Ahmad2020} propose \textit{MHWSMCB}, where various degrees are considered to choose influential nodes in a network as a influence maximization problem. The authors in \cite{Gupta2020} introduce an algorithm for overlapping community detection based on granular information of links and concepts of rough set theory.

\subsection{Other Approaches}

In \cite{Mendoza2014ExtractiveSS}, a mematic algorithm MA-SingleDocSum is introduced, the method uses evolutionary algorithm to solve the extractive summarization as a binary optimization problem. \cite{Yan2020} propose a hierarchical selective encoding network for both sentence-level and document-level representations and data containing important information is extracted. The method introduced in \cite{Batista2016} improves upon the cohesiveness of the summaries generated by extractive summarization systems. It is based on a post-processing step that binds dangling co-reference to the most important entity in a given co-reference chain. In \cite{Bonzanini2013} selectively removes unimportant sentences until a desired compression score is achieved. The work by \cite{Sonawane2019} model the document as semi-graph to extract both linear and non-linear relationships between the features. In \cite{parveen-etal-2015-topical}, a weighted graphical representation of the document is formed and coherence, non-redundance and importance are optimized using ILP (Inductive logic programming). \cite{Ozsoy2011} uses algorithms based on Latent Semantic Analysis to summarize Turkish and English text. \cite{Jang2021} proposes a deep-learning method to perform unsupervised summarization. The researchers in \cite{Nishino2013} use Langragian relaxation to solve summarization as combinatorial problem.

Several unsupervised algorithms have also been introduced to cluster the sentences and rank them. The methods use k-means \cite{Shetty2017} or fuzzy c-means \cite{ANam2018} due to their good generalization performance in other tasks also. Fuzzy c-means is not robust to noises and is sensitive to outliers in euclidean distance.

In \cite{Begun2009}, the method uses support vector machine (SVM) to train a summarizer using features like sentence position, sentence centrality, sentence similarity and several more. \cite{zopf-etal-2018-scores} uses sentence regression to score and greedily selects them to form the summary. In \cite{Chali2009}, the authors train an ensemble of SVMs over gram overlap, LCS, WLCS, skip-bigram, gloss overlap,BE overlap, length of sentence, position of the sentence, NE, cue word match, title match to approach the problem.

\section{Problem Statement}

Given a document $D$ consisting set of sentences $D = \{ s_1, s_2, s_3, \dots s_n \}$, where $n$ denotes the number of sentences in the document and $s_j$ is the \textit{j-th} sentence, $ 1\le j \le n$. The aim of of extractive summarization is to find a subset $D_s \subset D$ which contains different important topics mentioned in the complete document. It is expected to have $|D_s| \ll |D|$ where $| . |$ represent the number of sentences in the set.

A document consists of vast information covering various subtopics and a common main theme connecting them. Coverage means that the summary extracted by algorithm should cover most of the subtopics. Poor coverage of subtopics is indicated by absence of some relevant sentences. While extracting the sentences, deciding the importance based on relevance alone can be misleading and ignoring lesser covered but important subtopics. Therefore, focus of the algorithm on both relevance and coverage is necessary.

\section{Methodology} \label{sec:method}

\subsection{Graph Construction} \label{sec:graph-constr}

\begin{figure}[htbp]
    \centering
    \includegraphics[width=\textwidth]{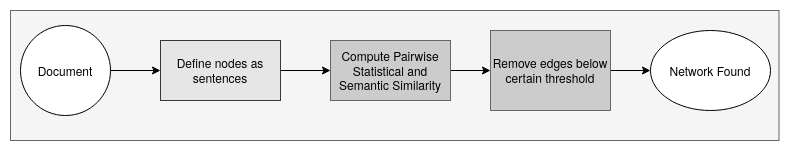}
    \caption{Flow diagram showing construction of graph-network from a given document.}
    \label{fig:graph_constr}
\end{figure}

The document is split into sentences. Let the graph formed be $G = (V, E)$ where elements in set $V$ are the sentences and $E$ represent the set of edges between a pair of sentences. The presence of an edge between a pair of sentences is decided by a weighted sum of their statistical and semantic similarities. A hyperparameter, $\delta_e$ is chosen and if a pair has similarity lesser than $\delta_e$, no edge exists between them. Clearly, a high value of $\delta_e$ encourages lesser number of edges and a lower value will include all the $\binom{V}{2} $ edges in the graph. An appropriate threshold will retain relations between important sentences and discard edges between insignificant sentences. This workflow is represented in the figure \ref{fig:graph_constr}

\subsection{Community Detection} \label{sec:comm-detect}

\begin{figure}[t]
    \centering
    \includegraphics[width=0.8\textwidth]{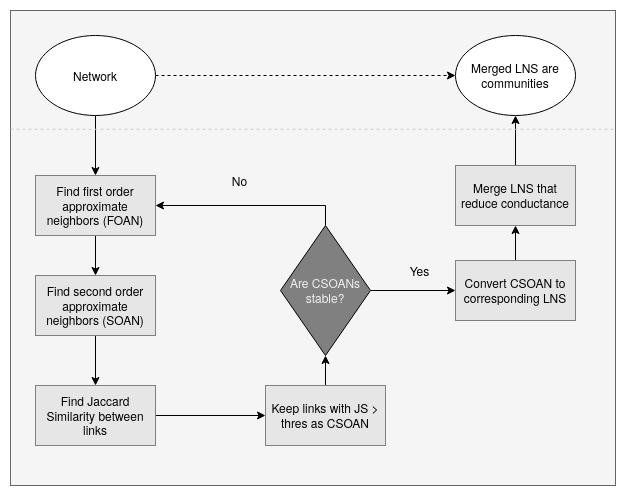}
    \caption{Flow diagram showing detection of communities by the algorithm.}
    \label{fig:comm_detect}
\end{figure}

The graph $G = (V, E)$ formed thus is a weighted graph, let $w_{i, j}$ represent weight of edge $e_{i, j}$ between node $v_i$ and node $v_j$. Additionally, it satisfies the condition: $w_{i, j} \ge \delta_e$ The figure \ref{fig:comm_detect} shows the flow diagram for this section of algorithm. The algorithm introduces terminoligies mentioned below:
\begin{enumerate}
    \item \textbf{FOAN: First Order Approximate Neighbors - } First Order Approximate Neighbors of a link $e_{i, j}$ are defined by :
        \begin{equation} \label{eqn:FOAN}
            F(e_{i, j}) = \{ e_{i, k} : k \in N_i \} \cup \{ e_{k, j} : k \in N_j \}
        \end{equation}
        where $N_i$ and $N_j$ represent the nodes connected to $v_i$ and $v_j$.

    \item \textbf{SOAN: Second Order Approximate Neighbors - } Second Order Approximate Neighbors of a link $e_{i, j}$ are defined by :
        \begin{equation} \label{eqn:SOAN}
            S(e_{i, j}) = \bigcup \{ F(e_{m, n}) : (m, n) \in F(e_{i, j}) \}
        \end{equation}
    
    \item \textbf{JS: Jaccard Similarity - } Jaccard Similarity between two vectors \textbf{x} and            \textbf{y} is given by:
        \begin{equation} \label{eqn:js}
            J_\mathcal{W}(\textbf{x}, \textbf{y}) = \frac{\sum_i min(x_i, y_i)}{\sum_i max(x_i, y_i)}
        \end{equation}
    
    \item \textbf{CSOAN: Constrained Second Order Approximate Neighbors - } Constrained Second Order Approximate Neighbors of a      link $e_{i, j}$ are defined by :
        \begin{equation} \label{eqn:cSOAN}
            C(e_{i, j}) = \{ e_{a, b} : e_{a, b} \in S(e_{i, j}) | J_{\mathcal{W}}(\overline{e}_{i, j}, \overline{e}_{a, b}) \ge \delta_{csoan} \}
        \end{equation}
    
    \item \textbf{LNS: Link Node Set - } Link Node Set of a link $e_{i, j}$ is defined by :
        \begin{equation} \label{eqn:lns}
            L(e_{i, j}) = \{ v_m, v_n : e_{m, n} \in C(e_{i, j}) \}
        \end{equation}
        So, $L(e_{i, j}) \subset V$ whereas $C(e_{i, j}) \subset E$, $S(e_{i, j}) \subset E$, and $F(e_{i, j}) \subset E$.
        
    \item \textbf{Conductance - } Conductance of a graph $G = (V, E)$ is given by :
        \begin{equation} \label{eqn:conductance}
            \phi(G) = \min_{S \in V; 0\le a(S) \le a(V)/2} \frac{\sum_{i \in S; j \in \overline{S} }a_{i, j}}{a(S)}
        \end{equation}
\end{enumerate}

The algorithm processes the graph of document through several steps iteratively unless a stable set of edges is obtained. A set of first order approximate neighbors is formed for every edge in graph, using the equation \ref{eqn:FOAN}. Next, a set of second order approximate neighbors is formed by the union of first order approximate neighbors of every edge in the first order approximate neighbors of the target edge (equation \ref{eqn:SOAN}). The cardinality of this set determines the number of nodes which share a strong similarity with the nodes of given edge. The sum of the weights of the elements in $S(e_{i, j})$ is higher for an edge with greater importance in the graph. Every set $F(e_{i, j}$ can be represented as vector with orthogonal components corresponding to the weight of edge elements. So every edge $e_{i, j}$ has a $S(e_{i, j})$ which can be expressed as a vector $\overline{e}_{i, j}$, given by equation \ref{eqn:set_to_vec}.

\begin{equation} \label{eqn:set_to_vec}
    \overline{e}_{i, j} = \sum_{t \in S(e_{i, j})} w_t \cdot \hat{t}
\end{equation}

For every pair of edges in the graph $e_{i, j}$ and $e_{m, n}$, Jaccard Similarity is calculated between their corresponding vectors $\overline{e}_{i, j}$ and $\overline{e}_{m, n}$ using equation \ref{eqn:js}. A higher coefficients for an edge indicate greater similarity and higher importance than the other edges. The coefficients found are used to filter out edges with low similarity in $S(e_{i, j})$. A threshold $\delta_{cson}$ is used to calculate the constrained second order approximate neighbors $C(e_{i, j})$ using equation \ref{eqn:cSOAN}. The set $C(e_{i, j})$ is used as $F(e_{i, j})$ in the next iterative step (if need be). The loop stops processing an edge $e_{i, j}$ when the $C(e_{i, j})$ for a step is same as $C(e_{i, j})$ in previous iteration, and the set $C(e_{i, j})$ is deemed stable.

When stable sets of $C(e_{i, j})$ for every edge $e_{i, j}$ are found out, loop completely terminates. Every set $C(e_{i, j})$ is used to make the corresponding link node set $L(e_{i, j})$ using equation \ref{eqn:lns}. Next, conductance of every $L(e_{i, j})$ is calculated using equation \ref{eqn:conductance}. A pair of $L(e_{i, j})$ and $L(e_{m, n})$ are merged if the resultant set $L(e_{i, j}) \cup L(e_{m, n})$ has a conductance lower than the individual sets. $L(e_{i, j})$ are merged (union) until conductance can no longer be reduced. The resultant set of node sets correspond to the communities detected.

\subsection{Finding Important Sentences}

Finding most significant sentences in document network is equivalent to finding most influential nodes in a network. Let the communities detected by following algorithm in section \ref{sec:comm-detect} are $\mathcal{H}$. The subgraphs formed using $G=(V, E)$ and nodes in $H_i \in H$ are overlapping in nature, hence the influence of a node in graph $G$ is determined by the influence in $H_i$ and also by $H_{j} \in \mathcal{H}, j\ne i$.

\begin{figure}[ht]
    \centering
    \includegraphics[width=0.9\textwidth]{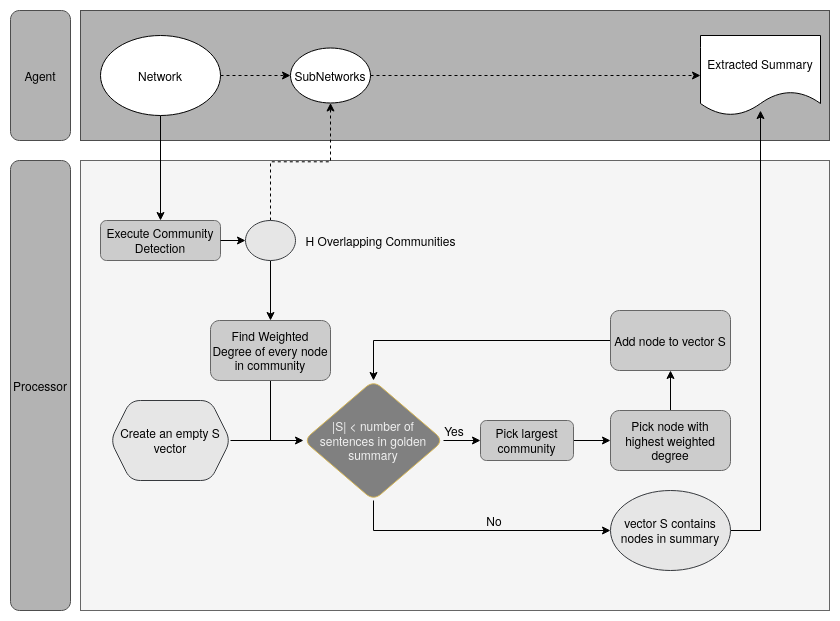}
    \caption{Flow diagram showing extraction of most important sentences in the document using propose CoRank method}
    \label{fig:infl_nodes}
\end{figure}

For every subgraph using nodes of $H_i \in \mathcal{H}$, weighted degree of each node is calculated. A larger weighted degree of a node signifies larger influence in the community. Additionally, a larger community is responsible for larger influence in the graph. So, the algorithm picks largest community and the node with largest weighted degree. This node is removed from the community and again the algorithm picks. The step mentioned is iteratively done until a desired number of nodes are extracted. The figure \ref{fig:infl_nodes} shows the flow diagram for this section of algorithm.
        where $N_i$ and $N_j$ represent the nodes connected to $v_i$ and $v_j$.

\section{Explanation of the proposed methodology using example network} \label{sec:example}

\begin{figure}[ht]
    \centering
    \includegraphics[width=0.5\textwidth]{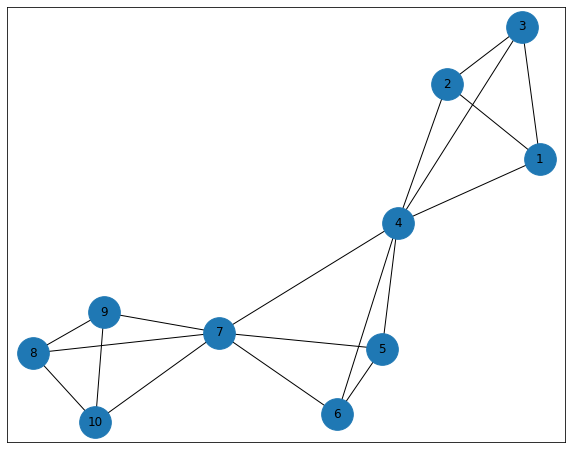}
    \caption{Flow diagram showing detection of communities by the algorithm.}
    \label{fig:example_graph}
\end{figure}

To further illustrate the process undertaken by the algorithm, a graph $G = (V, E)$ is taken as given in figure \ref{fig:example_graph}. To assign the weights to this example network, degree of the nodes is calculated. A parameter $\theta = \frac{E}{V}$ is used as threshold. For every link, if either of nodes have a degree greater than $\theta$, weight is drawn from a random number generator (between 2 and 10), available through the \textit{random} package in python 3, otherwise the degree is set to unity. The weight of each link for the example network thus found is mentioned in the table \ref{table:weights_example}. Table 2 presents the index mapping of the edges of the example network given in figure 4.

\begin{table}[ht]
    \centering
    \caption{Weights of the edges in the example graph(given in Fig.4)}
    \begin{tabular}{|l|l|l|l|l|l|l|l|} 
    \hline
    \textbf{Edge} & \textbf{Weight} & \textbf{Edge} & \textbf{Weight} & \textbf{Edge} & \textbf{Weight} & \textbf{Edge} & \textbf{Weight}  \\ 
    \hline
    (1, 2)        & 9                  & (1, 3)        & 6                  & (1, 4)        & 5                  & (2, 3)        & 7                   \\ 
    \hline
    (2, 4)        & 2                  & (3, 4)        & 5                  & (4, 5)        & 2                  & (4, 6)        & 2                   \\ 
    \hline
    (4, 7)        & 5                  & (5, 6)        & 2                 & (5, 7)        & 8                 & (6, 7)        & 5                  \\ 
    \hline
    (7, 8)        & 9                 & (7, 9)        & 3                 & (7, 10)       & 9                 & (8, 9)        & 2                  \\ 
    \hline
    (8, 10)       & 4                 & (9, 10)       & 6                 & -             & -                  & -             & -                   \\
    \hline
    \end{tabular}
    \label{table:weights_example}
\end{table}

\begin{table}
    \centering
    \caption{Index Mapping of the edges in the example graph(given in Fig.4)}
    \label{table:mapping_example}
    \resizebox{\columnwidth}{!}{%
    \begin{tabular}{|l|l|l|l|l|l|l|l|} 
    \hline
    \textbf{Edge} & \textbf{Map Index} & \textbf{Edge} & \textbf{Map Index} & \textbf{Edge} & \textbf{Map Index} & \textbf{Edge} & \textbf{Map Index}  \\ 
    \hline
    (1, 2)        & 1                  & (1, 3)        & 2                  & (1, 4)        & 3                  & (2, 3)        & 4                   \\ 
    \hline
    (2, 4)        & 5                  & (3, 4)        & 6                  & (4, 5)        & 7                  & (4, 6)        & 8                   \\ 
    \hline
    (4, 7)        & 9                  & (5, 6)        & 10                 & (5, 7)        & 11                 & (6, 7)        & 12                  \\ 
    \hline
    (7, 8)        & 13                 & (7, 9)        & 14                 & (7, 10)       & 15                 & (8, 9)        & 16                  \\ 
    \hline
    (8, 10)       & 17                 & (9, 10)       & 18                 & -             & -                  & -             & -                   \\
    \hline
    \end{tabular}%
    }
\end{table}

\subsection{Part A: Community Detection}

\subsubsection{Step 1: Find First Order Approximate Neighbors (FOAN)}
FOAN of every edge in the network is calculated using the equation \ref{eqn:FOAN}. The table \ref{tab:example_foan} shows the calculated FOAN for the example network given in Fig.4.

\begin{table}[htbp]
    \centering
    \begin{tabular}{c|c}
        Edge & Elements in FOAN (Edge Map Indices) \\
        \hline
        (1, 2) & {1, 2, 3, 4, 5} \\
        (1, 3) & {1, 2, 3, 4, 6} \\
        (1, 4) & {1, 2, 3, 5, 6, 7, 8, 9} \\
        (2, 3) & {1, 2, 4, 5, 6} \\
        (2, 4) & {1, 3, 4, 5, 6, 7, 8, 9} \\
        (3, 4) & {2, 3, 4, 5, 6, 7, 8, 9} \\
        (4, 5) & {3, 5, 6, 7, 8, 9, 10, 11} \\
        (4, 6) & {3, 5, 6, 7, 8, 9, 10, 12} \\
        (4, 7) & {3, 5, 6, 7, 8, 9, 11, 12, 13, 14, 15} \\
        (5, 6) & {7, 8, 10, 11, 12} \\
        (5, 7) & {7, 9, 10, 11, 12, 13, 14, 15} \\
        (6, 7) & {8, 9, 10, 11, 12, 13, 14, 15} \\
        (7, 8) & {9, 11, 12, 13, 14, 15, 16, 17} \\
        (7, 9) & {9, 11, 12, 13, 14, 15, 16, 18} \\
        (7, 10) & {9, 11, 12, 13, 14, 15, 17, 18} \\
        (8, 9) & {13, 14, 16, 17, 18} \\
        (8, 10) & {13, 15, 16, 17, 18} \\
        (9, 10) & {14, 15, 16, 17, 18}

    \end{tabular}
    \caption{FOAN of all edges in graph. The numbers shown in right column are the edge map indices for brevity, refer to table \ref{table:mapping_example}.}
    \label{tab:example_foan}
\end{table}

\subsubsection{Step 2: Find Second Order Approximate Neighbors (SOAN)}
SOAN of every edge in the network is calculated using the equation \ref{eqn:SOAN}. The table \ref{tab:example_soan} shows the calculated FOAN for the example network(given in Fig.4).

\begin{table}[htbp]
    \centering
    \begin{tabular}{c|c}
        Edge & Elements in SOAN (Edge Map Indices) \\
        \hline
        (1, 2) & {1, 2, 3, 4, 5} \\
        (1, 3) & {1, 2, 3, 4, 6} \\
        (1, 4) & {1, 2, 3, 5, 6, 7, 8, 9} \\
        (2, 3) & {1, 2, 4, 5, 6} \\
        (2, 4) & {1, 3, 4, 5, 6, 7, 8, 9} \\
        (3, 4) & {2, 3, 4, 5, 6, 7, 8, 9} \\
        (4, 5) & {3, 5, 6, 7, 8, 9, 10, 11} \\
        (4, 6) & {3, 5, 6, 7, 8, 9, 10, 12} \\
        (4, 7) & {3, 5, 6, 7, 8, 9, 11, 12, 13, 14, 15} \\
        (5, 6) & {7, 8, 10, 11, 12} \\
        (5, 7) & {7, 9, 10, 11, 12, 13, 14, 15} \\
        (6, 7) & {8, 9, 10, 11, 12, 13, 14, 15} \\
        (7, 8) & {9, 11, 12, 13, 14, 15, 16, 17} \\
        (7, 9) & {9, 11, 12, 13, 14, 15, 16, 18} \\
        (7, 10) & {9, 11, 12, 13, 14, 15, 17, 18} \\
        (8, 9) & {13, 14, 16, 17, 18} \\
        (8, 10) & {13, 15, 16, 17, 18} \\
        (9, 10) & {14, 15, 16, 17, 18}
    \end{tabular}
    \caption{SOAN of all edges in graph. The numbers shown in right column are the edge map indices for brevity, refer to table \ref{table:mapping_example}.}
    \label{tab:example_soan}
\end{table}

\subsubsection{Step 3: Find Jaccard Similarity}

For every pair of $F(e_{i, j})$ corresponding vectors $\overline{e}_{i, j}$ are calculated and Jaccard similarity is found out using the equation \ref{eqn:js}. The Jaccard Similarity for the given example graph(shown in Fig.4) is given in table \ref{tab:example_js}.

\begin{table}[ht]
    \centering
    \resizebox{\columnwidth}{!}{%
    \begin{tabular}{c|c|c|c|c|c|c|c|c|c|c|c|c|c|c|c|c|c|c}
            &    1 &    2 &    3 &    4 &    5 &    6 &    7 &    8 &    9 &   10 &   11 &   12 &   13 &   14 &   15 &   16 &   17 &   18 \\
            \hline
            1 &  $\infty$ & & & & & & & & & & & & & & & & \\
            2 &  0.79 & $\infty$ & & & & & & & & & & & & & & & \\
            3 &  0.51 & 0.58 & $\infty$ & & & & & & & & & & & & & & \\
            4 &  0.71 & 0.79 & 0.51 & $\infty$ & & & & & & & & & & & & &\\
            5 &  0.53 & 0.60 & 0.70 & 0.53 & $\infty$ & & & & & & & & & & & &\\
            6 &  0.47 & 0.53 & 0.63 & 0.47 & 0.65  & $\infty$ & & & & & & & & & & &\\
            7 &  0.13 & 0.19 & 0.46 & 0.13 & 0.45 & 0.48  & $\infty$ & & & & & & & & & &\\
            8 &  0.14 & 0.20 & 0.49 & 0.14 & 0.48 & 0.51 & 0.64  & $\infty$ & & & & & & & & &\\
            9 &  0.09 & 0.13 & 0.30 & 0.09 & 0.30 & 0.31 & 0.51 & 0.46  & $\infty$ & & & & & & & &\\
            10 & 0.00 & 0.00 & 0.08 & 0.00 & 0.08 & 0.08 & 0.39 & 0.31 & 0.30  & $\infty$ & & & & & & &\\
            11 & 0.00 & 0.00 & 0.10 & 0.00 & 0.10 & 0.10 & 0.30 & 0.25 & 0.72 & 0.38  & $\infty$ & & & & & &\\
            12 & 0.00 & 0.00 & 0.10 & 0.00 & 0.10 & 0.10 & 0.30 & 0.25 & 0.72 & 0.38 & 0.91  & $\infty$ &&&&&&\\
            13 & 0.00 & 0.00 & 0.07 & 0.00 & 0.06 & 0.07 & 0.21 & 0.16 & 0.64 & 0.25 & 0.80 & 0.80  & $\infty$ &&&&\\
            14 & 0.00 & 0.00 & 0.06 & 0.00 & 0.06 & 0.07 & 0.20 & 0.15 & 0.62 & 0.25 & 0.76 & 0.76 & 0.80  & $\infty$ &&&&\\
            15 & 0.00 & 0.00 & 0.06 & 0.00 & 0.06 & 0.06 & 0.19 & 0.15 & 0.60 & 0.24 & 0.74 & 0.74 & 0.84 & 0.88  & $\infty$ &&&\\
            16 & 0.00 & 0.00 & 0.00 & 0.00 & 0.00 & 0.00 & 0.00 & 0.00 & 0.18 & 0.00 & 0.22 & 0.22 & 0.35 & 0.39 & 0.43  & $\infty$ &&\\
            17 & 0.00 & 0.00 & 0.00 & 0.00 & 0.00 & 0.00 & 0.00 & 0.00 & 0.27 & 0.00 & 0.33 & 0.33 & 0.47 & 0.51 & 0.55 &0.64  & $\infty$ &\\
            18 & 0.00 & 0.00 & 0.00 & 0.00 & 0.00 & 0.00 & 0.00 & 0.00 & 0.18 & 0.00 & 0.22 & 0.22 & 0.35 & 0.39 & 0.43 &0.45 &0.64  & $\infty$ \\

    \end{tabular}%
    }
    \caption{Jaccard Similarity }
    \label{tab:example_js}
\end{table}

\subsubsection{Step 4: Find Constrained Second Order Approximate Neighbors}

The algorithm uses the Jaccard Similarity found in the previous step to eliminate weaker relations in SOAN. Using a threshold $\delta_{cson} = 0.5$ and $\alpha^{t=1}_{decay} = 0.8$, CSON (table \ref{tab:example_csoan} is found. This completes the first iteration. The CSON acts as FOAN for the next iteration. The loop stops when $CSON_{new}$ is same as $CSON_{prev}$.

\begin{table}[hp]
    \centering
    \begin{tabular}{c|c}
        Edge & Elements in CSOAN (Edge Map Indices) \\
        \hline
        (1, 2) & {1, 2, 3, 4, 5} \\
        (1, 3) & {1, 2, 3, 4, 5, 6} \\
        (1, 4) & {1, 2, 3, 4, 5, 6} \\
        (2, 3) & {1, 2, 3, 4, 5} \\
        (2, 4) & {1, 2, 3, 4, 5, 6} \\
        (3, 4) & {2, 3, 5, 6, 8} \\
        (4, 5) & {7, 8, 9} \\
        (4, 6) & {6, 7, 8} \\
        (4, 7) & {7, 9, 11, 12, 13, 14, 15} \\
        (5, 6) & {10} \\
        (5, 7) & {9, 11, 12, 13, 14, 15} \\
        (6, 7) & {9, 11, 12, 13, 14, 15} \\
        (7, 8) & {9, 11, 12, 13, 14, 15} \\
        (7, 9) & {9, 11, 12, 13, 14, 15, 17} \\
        (7, 10) & {9, 11, 12, 13, 14, 15, 17} \\
        (8, 9) & {16, 17} \\
        (8, 10) & {14, 15, 16, 17, 18} \\
        (9, 10) & {17, 18} \\
    \end{tabular}
    \caption{CSOAN of all edges in graph. The numbers shown in right column are the edge map indices for brevity, refer to table \ref{table:mapping_example}.}
    \label{tab:example_csoan}
\end{table}

\subsubsection{Step 5: Merging Link Node Sets (LNS)} \label{sec:find_comms}

After 4 iterations, a stable set of CSON is formed. The CSOAN of each edge does not change in any further iteration. CSON are converted into LNS - Link Node Set using equation \ref{eqn:lns}. For every set of LNS, conductance is calculated and two LNS are merged (union) if and only if the resultant set has a lower conductance than the individual values. The values of stable CSON, LNS and their conductance is given in table \ref{tab:example_end_csoan}.

\begin{table}[htbp]
    \centering
    \resizebox{\columnwidth}{!}{%
    \begin{tabular}{c|c|c|c}
        Edge & Elements in CSOAN (Edge Map Indices) & Link Node Set & Conductance $\phi$\\
        \hline
        (1, 2) & {1, 2, 3, 4, 5, 6} & {3, 4, 1, 2} & 0.2\\
        (1, 3) & {1, 2, 3, 4, 5, 6} & {3, 4, 1, 2} & 0.2\\
        (1, 4) & {1, 2, 3, 4, 5, 6} & {3, 4, 1, 2} & 0.2\\
        (2, 3) & {1, 2, 3, 4, 5, 6} & {3, 4, 1, 2} & 0.2\\
        (2, 4) & {1, 2, 3, 4, 5, 6} & {3, 4, 1, 2} & 0.2\\
        (3, 4) & {1, 2, 3, 4, 5, 6} & {3, 4, 1, 2} & 0.2\\
        (4, 5) & {7} & {4, 5} & 0.778\\
        (4, 6) & {8} & {4, 6} & 0.778\\
        (4, 7) & {9, 11, 12, 13, 14, 15} & {8, 6, 5, 10, 7, 4, 9} & 0.334\\
        (5, 6) & {10} & {6, 5} & 0.667\\
        (5, 7) & {9, 11, 12, 13, 14, 15} & {8, 6, 5, 10, 7, 4, 9} & 0.334\\
        (6, 7) & {9, 11, 12, 13, 14, 15} & {8, 6, 5, 10, 7, 4, 9} & 0.334\\
        (7, 8) & {9, 11, 12, 13, 14, 15} & {8, 6, 5, 10, 7, 4, 9} & 0.334\\
        (7, 9) & {9, 11, 12, 13, 14, 15} & {8, 6, 5, 10, 7, 4, 9} & 0.334\\
        (7, 10) & {9, 11, 12, 13, 14, 15} & {8, 6, 5, 10, 7, 4, 9} & 0.334\\
        (8, 9) & {16} & {8, 9} & 0.667\\
        (8, 10) & {17, 18} & {8, 10, 9} & 0.334\\
        (9, 10) & {17, 18} & {8, 10, 9} & 0.334\\
    \end{tabular}%
    }
    \caption{CSOAN of all edges in graph after the loop stops. The numbers shown in right column are the edge map indices for brevity, refer to table \ref{table:mapping_example}.}
    \label{tab:example_end_csoan}
\end{table}

 After merging the LNS, the communities thus found are given in table \ref{tab:ex_comm_detect}.
\begin{table}[htbp]
\centering
\begin{tabular}{l|l|l|l|l}
\multicolumn{5}{c}{\textbf{Communities Detected}} \\
\hline
1, 2, 3, 4 & 1, 2, 3, 4, 5, 6 & 4, 5, 6, 7, 8, 9, 10 & 1, 2, 3, 4, 8, 9, 10 & 8, 9, 10 \\
\end{tabular}
\caption{Communities Detected}
\label{tab:ex_comm_detect}
\end{table}

\subsection{Finding Influential Nodes / Important Sentences}

After steps in section \ref{sec:find_comms}, a list of nodes in different communities is obtained. The overlapping communities are sorted in their decreasing length and the largest community is picked. In the given example $H = \{1, 2, 3, 4, 8, 9, 10\}$ is largest. The weighted degree of the nodes are found out as given in table \ref{tab:weighted degree}. Node $1$ has highest degree, it is removed from the community and added to a vector $S$. $S$ now contains $S = {1}$. Again the largest community is picked and found out to be $ H = \{4, 5, 6, 7, 8, 9, 10\}$, node $10$ has highest weighted degree of $7$, it is added to vector $S$. $S now contains {1, 7}$. Similarly, the operation is carried out till 5 nodes are found out. The resultant vector $S = \{1, 7, 3, 10, 2\}$. These are the most influential nodes in the network.

\begin{table}[htbp]
    \centering
    \begin{tabular}{c|c|c|c|c|c}
        Node & Weighted Degree & Node & Weighted Degree & Node & Weighted Degree \\ 
        \hline
        1 & 20 &
        2 & 18 &
        3 & 18 \\
        4 & 12 &
        8 & 6 &
        9 & 8 \\
        && 10 & 10 
    \end{tabular}
    \caption{Weighted degree of the nodes in community $H = \{1, 2, 3, 4, 8, 9, 10\}$}
    \label{tab:weighted degree}
\end{table}
\subsection{Complexity of proposed CoRank} \label{sec:complex}

For a graph $G = (V, E)$ where $V$ is the set of vertices and $E$ is the set of edges, complexity is computed of the proposed CoRank algorithm.

\begin{enumerate}
    \item \textbf{Community Detection -}
        \begin{enumerate}
            \item Computing FOAN: $O(|E|)$
            \item Computing SOAN: $O(|E|^2)$
            \item Finding Jaccard Similarity: $O(|E|^2 \cdot log(\overline{E}))$ where $\overline{E}$ represent average number of edges in FOAN.
            \item Computing CSON: $O(|E|^2)$
            \item Merging LNS: $O(|E| \cdot log(m))$ where $m$ is the average number of LNS merged.
        \end{enumerate}
        This results in complexity of $O(|E|) + N_{iter} \cdot (O(|E|^2) + O(|E|^2 \cdot log(\overline{E})) + O(|E| \cdot log(m))$ where $N_{iter}$ is the number of iterations performed to obtain a stable CSOAN, equivalent to $N_{iter} O(|E|^2 \cdot log(\overline{E}))$. The maximum number of iterations the algorithm will be at most $|E|$. This gives a
        upper bound on the complexity of this step as $O(|E|^3)$.
    
    \item \textbf{Influential Nodes - } For $g$ number of sentences in the golden summary, this step takes $g \cdot O(|V|^2)$. This is a poor upper bound as the number of edges formed during graph construction are far fewer than $\binom{|V|}{2}$ and this step takes very few seconds on real test cases.
    
    \item Thus, the complexity of proposed CoRank is $O(|E|^3)+O(|V|^2) $
\end{enumerate}
\subsection{Real Examples} \label{ss:ex_extracted}

Given a document with text as given in figure \ref{qt:example_doc}, the algorithm is used to perform the extractive summarization. It detects $|\mathcal{H}| = 23$ communities and the sentences mentioned in quotes below are extracted. ROUGE-n gram score is computed to analyse the efficacy of the extracted summary(a brief discussion about ROUGE is given in Section 6.3). The results obtained in terms of recall, precision, and f-score demonstrate the efficiency of the proposed approach.

\begin{quote} \label{qt:example_doc}
    \textbf{Document:} \\
    Source: \href{https://edition.cnn.com/2014/06/12/health/virus-chikungunya/index.html}{CNN News Article by Val Willingham} \\
    A debilitating, mosquito-borne virus called chikungunya has made its way to North Carolina, health officials say. It's the state's first reported case of the virus.
The patient was likely infected in the Caribbean, according to the Forsyth County Department of Public Health. Chikungunya is primarily found in Africa, East Asia and the Caribbean islands, but the Centers for Disease Control and Prevention has been watching the virus,+ for fear that it could take hold in the United States -- much like West Nile did more than a decade ago.
The virus, which can cause joint pain and arthritis-like symptoms, has been on the U.S. public health radar for some time. About 25 to 28 infected travelers bring it to the United States each year, said Roger Nasci, chief of the CDC's Arboviral Disease Branch in the Division of Vector-Borne Diseases.
"We haven't had any locally transmitted cases in the U.S. thus far," Nasci said.
But a major outbreak in the Caribbean this year -- with more than 100,000 cases reported -- has health officials concerned. Experts say American tourists are bringing chikungunya back home, and it's just a matter of time before it starts to spread within the United States.
Study: Beer drinkers attract mosquitoes

Study: Beer drinkers attract mosquitoes 01:26
After all, the Caribbean is a popular one with American tourists, and summer is fast approaching.
"So far this year we've recorded eight travel-associated cases, and seven of them have come from countries in the Caribbean where we know the virus is being transmitted," Nasci said.
Other states have also reported cases of chikungunya. The Tennessee Department of Health said the state has had multiple cases of the virus in people who have traveled to the Caribbean.
The virus is not deadly, but it can be painful, with symptoms lasting for weeks. Those with weak immune systems, such as the elderly, are more likely to suffer from the virus' side effects than those who are healthier.
The good news, said Dr. William Shaffner, an infectious disease expert with Vanderbilt University in Nashville, is that the United States is more sophisticated when it comes to controlling mosquitoes than many other nations.
"We live in a largely air-conditioned environment, and we have a lot of screening (window screens, porch screens)," Shaffner said. "So we can separate the humans from the mosquito population, but we cannot be completely be isolated."
Chikungunya was originally identified in East Africa in the 1950s. The ecological makeup of the United States supports the spread of an illness such as this, especially in the tropical areas of Florida and other Southern states, according to the CDC.
The other concern is the type of mosquito that carries the illness. Unlike most mosquitoes that breed and prosper outside from dusk to dawn, the chikungunya virus is most often spread to people by Aedes aegypti and Aedes albopictus mosquitoes.
These are the same mosquitoes that transmit the virus that causes dengue fever. They bite mostly during the daytime. The disease is transmitted from mosquito to human, human to mosquito and so forth. A female mosquito of this type lives three to four weeks and can bite someone every three to four days.
\end{quote}

\begin{quote} 
    \textbf{Extracted:}
    Chikungunya is primarily found in Africa, East Asia and the Caribbean islands, but the Centers for Disease Control and Prevention has been watching the virus,+ for fear that it could take hold in the United States much like West Nile did more than a decade ago. Experts say American tourists are bringing chikungunya back home, and it's just a matter of time before it starts to spread within the United States. The virus, which can cause joint pain and arthritis-like symptoms, has been on the U.S. public health radar for some time. \\
    \textbf{Golden Summary:}
    North Carolina reports first case of mosquito-borne virus called chikungunya. Chikungunya is primarily found in Africa, East Asia and the Caribbean islands. Virus is not deadly, but it can be painful, with symptoms lasting for weeks. 
\end{quote}

\begin{table}[htbp]
    \centering
    \begin{tabular}{l|lll}
         & \textbf{Recall} & \textbf{Precision} & \textbf{F1-score}  \\
        \hline
        \textbf{ROUGE-1} & 0.824034 & 0.64646465 & 0.72452830 \\
        \textbf{ROUGE-2}  & 0.517241 & 0.40540541 & 0.45454545 \\
        \textbf{ROUGE-3}  & 0.246753 & 0.19322034 & 0.21673004 \\
    \end{tabular}
    \caption{ROUGE-1, 2, 3 score on the given real example in Subsection \ref{ss:ex_extracted}}
\end{table}

\section{Analysis of Experiments} \label{sec:results}
To validate the efficiency of the proposed approach, experiments are conducted on the data sets given by Document Understanding Conference (DUC). DUC is a method assessment competition that allows researchers to assess the efficiency of various summarization methods on similar data sets. 

\begin{table}[bp]
  \caption{Statistics of Data sets}
  \label{tab:freq}
  \begin{tabular}{ccl}
    \toprule
     & DUC 2001 & DUC 2002\\
    \midrule
    Count of clusters & 30 & 59\\
    Length of summary & 100 words & 100 words\\
    Source of data & TREC-9 & TREC-9\\
    Count of documents & 309 & 567 \\
  \bottomrule
\end{tabular}
\end{table}
\subsection{Data sets}
A brief description about the statistics of the data set is presented in Table 10. DUC 2001 and DUC 2002 data set is used for experimental analysis and a comparative study is made for the system generated summaries and golden summaries. 
\subsection{Pre-processing}
Some linguistic techniques such as stemming, upper case removal, removal of stop words,  and segmentation of sentences has been used during pre-processing phase of the document in this experiment. The textual content of the document is divided into sentences during the process of segmentation. Words appearing frequently such as a, an, the etc. within the text are removed during stop word removal process, since they are considered irrelevant.  The process of stemming involves reduction of a word to its root stem. PorterStemmer is used for the stemming of word. The process of pre-processing is performed before the execution of the algorithm. 
\subsection{Evaluation Metric}
DUC has adopted ROUGE metric \cite{LinC} for evaluation of automatically generated summary, hence, the proposed research uses this metric for performance analysis of the proposed research. The quality of summary is measured using ROUGE metric, that basically counts the number of overlapping units such as word pairs, word sequences, and n-grams between the reference summary and candidate summary. ROUGE-1 and ROUGE-2 recall score is used in this research for the evaluation of automatic summaries. 
\subsection{Evaluation of performance}

\begin{table}
  \centering
  \begin{tabular}{|p{2cm}|c|c|c|c|c|c|c|c|}
    \hline
    \textbf{Methods} & \multicolumn{4}{c|}{\textbf{DUC01}} & \multicolumn{4}{c|}{\textbf{DUC02}}\\
    \cline{2-9}
    & \textbf{R-1} & \textbf{Rank} & \textbf{R-2} & \textbf{Rank} & \textbf{R-1} & \textbf{Rank} & \textbf{R-2} & \textbf{Rank} \\
    \hline
    FEOM & 0.4773 & 1 & 0.1855 & 5 & 0.4658 & 7 & 0.1249 & 8 \\ \hline
    CoRank & 0.4725 & 2 & 0.2011 & 2 & 0.4906 & 1 & 0.2306 & 1 \\ \hline
    NetSum & 0.4643 & 3 & 0.1770 & 7 & 0.4496 & 8 & 0.1117 & 9 \\ \hline
    CRF & 0.4551 & 4 & 0.1773 & 9 & 0.4401 & 10 & 0.1092 & 10 \\ \hline
    ESDS & 0.4540 & 5 & 0.1957 & 4 & 0.4790 & 5 & 0.2214 & 4 \\ \hline
    UnifiedRank & 0.4538 & 6 & 0.1765 & 8 & 0.4849 & 2 & 0.2146 & 5 \\ \hline
    MA & 0.4486 & 7 & 0.2014 & 1 & 0.4828 & 3 & 0.2284 & 3 \\ \hline
    QCS & 0.4485 & 8 & 0.1852 & 6 & 0.4487 & 9 & 0.1877 & 7 \\ \hline
    LexRank & 0.4468 & 9 & 0.1989 & 3 & 0.4796 & 4 & 0.2295 & 2 \\ \hline
    SVM & 0.4463 & 10 & 0.1702 & 10 & 0.4324 & 11 & 0.1087 & 11 \\ \hline
    CollabSum & 0.4404 & 11 & 0.1623 & 12 & 0.4719 & 6 & 0.2010 & 6 \\ \hline
    ManifoldRanking & 0.4336 & 12 & 0.1664 & 11 & 0.4233 & 12 & 0.1068 & 12 \\ \hline
    DPSO & 0.3993 & 13 & 0.0832 & 13 & 0.4172 & 13 & 0.1026 & 13 \\ \hline
    0-1 non-linear & 0.0.3876 & 14 & 0.0778 & 14 & 0.4097 & 14 & 0.0937 & 14 \\ \hline
    
  \end{tabular}
  \caption{R-1 and R-2 recall score for DUC01 and DUC02 dataset}
\end{table}

This section deals with a comparative analysis of the performance of the proposed approach with some recent works. The proposed CoRank method is compared with some baseline methods: (a) ESDS \cite{Mendoza2015} – a search algorithm based on binary optimization, (b) manifold ranking \cite{Wan2007}- greedy search involving probabilistic approach, (c) NetSum \cite{Svore} - approach based on neural networks, (d) MA  \cite{Mendoza2014}- local search and genetic operators based metaheuristic approach, (e) CRF \cite{Shen}- approach based on conditional random field, (f) FEOM \cite{Song} – evolutionary algorithm involving fuzzy approach, (g) SVM \cite{Yeh2015} - mathematical approach, (h) QC \cite{Dunlavy} - hidden markov model based approach, (i)  0–1 non linear \cite{Alguliev2013} -  evolutionary algorithm approach based on binary PSO, (j) UnifiedRank \cite{wan-2010-towards} - ranking approach based on graph, (k) CollabSum \cite{Wan2007a} - clustering approach along with ranking of graphs, (l)  LexRank \cite{Erkan2004} - method involving ranking of graphs, (m) DPSO \cite{Alguliev2011a} - optimization approach based on evolutionary algorithm. The above methods have accomplished good results on DUC01 and DUC02 data sets, due to this reason they are selected for comarison with the proposed CoRank method.

Table 11 presents R-1 and R-2 recall score for DUC01 and DUC02 data sets. As is concerned for DUC01 data set, FEOM and CoRank acheive first and second place for R-1 recall score, whereas MA and CoRank  accomplish first and second place for R-2 recall score. For DUC02 data set, Corank is performing best for both R-1 and R-2 recall score, with DPSO and 0-1 non-linear performing worst in every case.
\begin{table}
  \centering
  \begin{tabular}{|p{2cm}|c|c|c|c|}
    \hline
    \textbf{Methods} & \multicolumn{2}{c|}{\textbf{DUC01}} & \multicolumn{2}{c|}{\textbf{DUC02}}\\
    \cline{2-5}
    & \textbf{R-1} &  \textbf{R-2} & \textbf{R-1} &  \textbf{R-2}  \\
    \hline
    FEOM & (-)0.96  & (+) 8.46 & (+) 5.37 & (+) 84.87  \\ \hline
    NetSum & (+)1.81  & (+) 13.67 & (+) 9.16 & (+) 106.71 \\ \hline
    CRF & (+)3.87  & (+) 16.10 & (+) 11.52 & (+) 111.45 \\ \hline
    ESDS & (+)4.12  & (+) 2.81 & (+) 2.46 & (+) 4.29 \\ \hline
    UnifiedRank & (+)4.16  & (+) 13.99 & (+) 1.22 & (+) 7.60 \\ \hline
    MA & (+)5.37  & (-) 0.10 & (+) 1.66 & (+) 1.09 \\ \hline
    QCS & (+)5.40  & (+) 8.64 & (+) 9.38 & (+) 23.02 \\ \hline
    LexRank & (+)5.80  & (+) 1.16 & (+) 2.34 & (+) 0.61 \\ \hline
    SVM & (+)5.92  & (+) 18.21 & (+) 13.51 & (+) 112.42 \\ \hline
    CollabSum & (+)7.33  & (+) 23.97 & (+) 4.01 & (+) 14.88 \\ \hline
    ManifoldRanking & (+)9.02  & (+) 20.91 & (+) 15.95 & (+) 116.20 \\ \hline
    DPSO & (+)18.38  & (+) 141.83 & (+) 17.64 & (+) 125.05 \\ \hline
    0-1 non-linear & (+)21.96  & (+) 158.61 & (+) 19.79 & (+) 146.42 \\ \hline
    
  \end{tabular}
  \caption{Relative improvement of CoRank with other methods}
\end{table}
A significant improvement was observed in the performance of the proposed approach for R-1 and R-2 recall score in comparison with other methods. The relative improvement of CoRank wrt other approaches is presented in Table 12. As is evident, CoRank outperforms state-of-the-art methods and achieves highest R-1 and R-2 recall score for DUC01 and DUC02 data sets. Relative improvement is used as a measure for comparison. The formula for calculating relative improvement is (c-b)*100/b, where a comparison of c is made with a. For showing the relative improvement of CoRank with other methods “+” sign is used, whereas “-” means opposite. As shown in Table 12, CoRank has outperformed other methods for R-1 and R-2 recall score on DCU01 and DUC02 data sets. For DUC01 data set, only FEOM and MA have performed better than CoRank. FEOM has shown an improvement of 0.96\% for R-1 metric whereas, MA is performing 0.10\% better than CoRank for R-2 metric. CoRank has outperformed state-of-the-art methods for DUC02 dataset. The reason for CoRank's good performance is, it is clustering sentences based on large sub-topics that are carrying more weight, and obtained salient sentences have maximum diversity and coverage.

Based on Table 12 observation, we can draw following conclusion:
\begin{itemize}
\item  The proposed approach CoRank has outperformed state-of-the-art-methods for DUC02 data set and performed competitively well for DUC01 data set except for DE and MA method.
\item Although LexRank and UnifiedRank are graph based approaches, but its performance is less in comparison to CoRank, which is a combination of clustering and graph ranking method.
\item CoRank, FEOM,  UnifiedRank,  ESDS,  LexRank, and MA are unsupervised methods that have performed better than SVM , a supervised approach.
\item 0-1 non-linear and DPSO have failed to perform, since they don’t use clustering concept.
    
\item Following conclusion is drawn when CoRank is compared with FEOM, MA, and NetSum, the top performing methods:(a) combination of graph based approaches with clustering has a bright future for further research in the domain of extractive summarization.
\end{itemize}

\section{Conclusion} \label{sec:conclusion}
This research work proposes CoRank: a clustering combined graph ranking approach for generating extractive summaries. It aims to cover two aspects: (a) diversity – obtained summary should not cover redundant information; (b) coverage – resultant summary should contain different main topics , sub-topics of the oringinal source document. Initially, a clustering algorithm is proposed that groups sentences into clusters based on topics and sub-topics. Then, from every cluster, most salient and representative sentences are selected using proposed CoRank algorithm. 

The preformance of the CoRank algorithm is validated on DUC01 and DUC02 datasets in terms of Recall-1 and Recall-2 measure. The proposed approach obtains best results for DUC02 dataset beating the best performing MA approach by 1.66\% and 1.09\% . It obtains good results for DUC01 dataset, however, slightly lagging behind FEOM approach. Other graph based approaches such as LexRank and UnifiedRanking are not so effective since they lack the concept of clustering. The reason for the promising results of the proposed CoRank approach is, it is able to extract main topics and sub-topics from the main text with maximum coverage and diversity. 

The future work remains to include more techniques based on optimization, and use different combination of similarity measures for extracting summaries. 

\bibliographystyle{ACM-Reference-Format}
\bibliography{main}
\end{document}